\documentclass[litterpaper,12pt]{article}
\usepackage{multicol}
\usepackage[nomarkers,figuresonly,nofiglist]{endfloat}
\usepackage[dvipsnames]{xcolor}
\usepackage{pdflscape}
\usepackage{blindtext}
\usepackage{cite}
\usepackage{amsmath}
\usepackage{bm}
\usepackage{amsbsy}
\usepackage{amssymb}
\usepackage{verbatim}
\usepackage{subcaption}
\usepackage{array}
\newcolumntype{P}[1]{>{\centering\arraybackslash}p{#1}}
\newcolumntype{M}[1]{>{\centering\arraybackslash}m{#1}}
\usepackage{booktabs}
\usepackage{longtable}
\usepackage{caption}
\usepackage{tabularx}
\usepackage[english]{babel}
\usepackage{multirow}
\usepackage{pdflscape}
\usepackage{epstopdf}
\usepackage[top=1in, bottom=1in, left=1in, right=1in]{geometry}
\usepackage{authblk}
\ifx\pdfoutput\undefined
\usepackage[dvips]{graphicx}
\else
\usepackage[pdftex]{graphicx}
\fi

\usepackage{graphicx}
\usepackage{dcolumn}
\usepackage{bm}
\usepackage{amsfonts}
\usepackage{booktabs}
\usepackage{siunitx}
\usepackage{amsmath,amssymb}
\usepackage{fdsymbol}
\usepackage{xcolor}
\usepackage{pifont}
\usepackage{bbding}

\title{Breath Figure Spot: a Recovery Concentration manifestation \vspace{2em}}

\author[a,b,+]{Ali Alshehri}
\author[a]{Pirouz Kavehpour}

\affil[a]{Mechanical and Aerospace Engineering Department, Henry Samueli School of Engineering and Applied Science, University of California, Los Angeles, CA 90095, USA}
\affil[b]{Mechanical Engineering Department, King Fahd University of Petroleum and Minerals (KFUPM), Dhahran 31261, Saudi Arabia}
\affil[+]{Corresponding Author: Phone: +1 (310)-849-7913 \protect\\
E-mail: aalshehri@ucla.edu \vspace{3em}} 

\newpage

\begin{document}

\maketitle
\newpage

\begin{abstract}
Directing a jet of humid air to impinge on a surface that is cooled below the dew point results in micro-sized water droplets. Lord Rayleigh discussed the phenomenon by contrasting clean to flame-exposed glass and called such behaviour Breath Figures (BF). Historically, utilizing dew as a water source was investigated by several scientists dating back to Aristotle. However, due to the degrading effects of air as a non-condensable gas (NCG) such efforts are limited to small scale water production systems and exhaled breath condensate (EBC) technology, to name a few. Recently, the concept of BF has been utilized extensively in the generation of micro-scale polymer patterns as a self-assembly process. However, the generation of BF on surfaces while being impinged by a humid air jet has not been quantified. In this work, we illustrate that a BF spot generated on a cooled surface is a manifestation of a recovery concentration. The concept is analogous to the concept of adiabatic-wall temperature defined for heat transfer applications. Upon closer examination of the vapor concentration distribution on a cooled impinged surface, we found that the distribution exhibits distinct regimes depending on the radial location from the center of the impingement region. The first regime is confined within the impingement region, whereas the second regime lies beyond this radial location including the wall jet region. Scaling analysis as well as numerical solution of the former regime shows that the maximum concentration on the surface is equivalent to its counterpart of a free unbounded jet with similar geometrical conditions. Additionally, the scaling analysis of the latter regime reveals that the jet speed and standoff distance are not important in determining the recovery concentration. However, the recovery concentration is found to vary monotonically with the radial location. Our conclusions are of great importance in optimizing jet impingement where condensation phase change is prevalent.
\end{abstract}

\textbf{Keywords: Condensation; Phase Change; Mass Transfer, Breath Figure }

\newpage

\section*{NOMENCLATURE}

\begin{tabbing}
	\hspace*{3.5 cm}\=\kill
$a_v$ \> constant associated with the Gaussian distribution of the jet velocity \\ 
$a_\omega$ \> constant associated with the Gaussian distribution of the jet concentration \\ 
$C_{1-4}$ \> constants that are independent of jet velocity and standoff distance \\ 
$D$ \> Tube diameter, m \\ 
$g$ \> Gravitational acceleration, m/s$^2$ \\ 
$H$ \> Standoff distance between tube exit and condensation surface, m \\
$L$ \> Tube length, m \\ 
$D$ \> Tube diameter, m \\ 
$M$ \> Molecular weight, kg/mol \\ 
$P$ \> Pressure, Pa \\ 
$Q$ \> Volumetric flow rate, m$^3$/s \\ 
$Re$ \> Reynolds number \\ 
$RH$ \> Relative humidity \\ 
$D$ \> Tube diameter, m \\ 
$r$ \> Radial direction in polar coordinate system \\ 
$T$ \> Temperature , $^o$C \\
$v$ \> velocity, m/s \\ 
$z$ \> Axial direction in polar coordinate system \\

\textbf{Greek symbols}\\
$\nu$ \> Kinematic viscosity, m$^2$/s\\
$\rho$ \> Density, kg/m$^3$  \\
$\omega$ \> Species mass fraction defined as the ratio between the density of \\ \> a certain species to the total density at a prescribed state \\
$\delta$ \> Thickness of the boundary layer in the wall jet region, m \\

\textbf{Superscripts and subscripts}\\
$a$ \> Property related to air  \\
$BF$ \> Related to the diameter of the Breath Figure spot \\
$j$ \>  Property related to the jet at the tube exit condition \\
$max$ \> Property related to the maximum value in a velocity/concentration \\ \> distribution, usually located at the center (r=0)  \\
$o$ \> Related to the characteristic velocity \\
$r$ \> Property related to the recovery concentration which is defined as the \\ \> concentration on the wall where no condensation is present \\
$s$ \> Property related to condensation surface condition \\
$v$ \> Property related to the water vapor  \\
$\infty$ \>  Evaluated at the ambient conditions\\

\end{tabbing}

\newpage

\section{Introduction }

Condensation is a prevalent phenomenon in nature and industry, yet not fully explored. In nature, most living species rely evidently on condensed atmospheric vapor. Moreover, some plants and animals get their share of fresh water by evolutionary modified surfaces that enhance the condensation process, such as the Darkling beetles \cite{seely1979}, and Sequoia Sempervirens \cite{simonin2009}. Utilizing the phenomenon in numerous applications has been a course of scientific curiosity for a very long time, dating back to Aristotle (300 BC). In modern era, utilizing condensation has gone beyond large scale desalination plants to micro- and nano-scale lithography techniques \cite{jadhav2012,zhang2015breath}. In daily experience, people observe that upon breathing against a glass surface, white traces of condensate are generated. Upon a closer look under the microscope, such traces are composed of sessile droplets of a micron size \cite{beysens1986,beysens1986growth,beysens1991,fritter1991experiments,beysens2018dew}. External lighting scatters in all directions from dewed surfaces, therefore, they appear cloudy. However, old observations by Aitken \cite{aitken1895,aitkek1911,aitken1913} and Lord Rayleigh \cite{rayleigh1911,rayleigh1912} discussed that flame-exposed glass does not show such cloudiness. In their discussion, they termed such behaviour as \textit{breath figures} (BF) for obvious reasons. It is with our present understanding of surface energy effect that we are aware of wettability importance. Today, the phenomenon has been utilized in self assembly processes to produce honeycomb polymer patterns \cite{yabu2018fabrication,zhang2015breath,park2004breath,stenzel2006formation}. 


The presence of untraceable amounts of Non-condensable gases (NCG), such as air, in condensation processes has shown to dramatically reduce the condenser efficiency \cite{othmer1929,huang2015}. The reason of this reduction is the accumulation of NCG on the liquid-vapor interface introducing a layer that is NCG-rich. The condensation rate becomes solely limited by the diffusion of vapor through this layer. Researchers have shown that heat transfer is thus limited by this layer's thermal resistance. Even though experimental studies have been successful in reducing NCG effect by means of vacuuming test chambers to environment \cite{rose2015}, it is a highly impractical solution in large scale equipment. NCG can break through equipment via leak points, which is a problem of its own, or as a chemical reaction product of vapor interacting with the equipment material \cite{Zhang2017}. 

In efforts to mitigate the negative effect of NCG, other active techniques have been utilized, such as extended surfaces \cite{chafik2004,chafik2003,farid2003,chang2014experimental}; direct contact between gas and cooling medium \cite{klausner2006,dawoud2006,hu2011,agboola2015,tow2014,liu2016,sadeghpour2019water}; and different NCG carriers \cite{narayan2011,arabi2003}. Even though the former two solutions are promising, the latter seems to address the problem at its core, i.e. the effect of vapor diffusion through the diffusion layer which in result affects the heat and mass transfer. However, improvements from those techniques come with great material cost (former two techniques) or industrial impracticality (latter technique). Investigating the problem of NCG further shows that the solution lies within two possibilities; (1) increasing heat/mass transfer contact area ($A$); (2) increasing heat/mass transfer coefficient ($h$). The optimal solution should be obtained by maximizing the design parameter ($hA$) while minimizing the required cost. Jet impingement of heat transfer fluids has shown a great potential in increasing the heat transfer coefficient for single phase \cite{martin1977,jambunathan1992,viskanta1993,lienhard1995,lienhard2006,webb1995single} as well as multi-phase applications \cite{xiao2010drying,supmoon2013influence,sarkar2004fluid,francis1996jet,santos2015study,mujumdar201415,polat1993heat}.


Utilization of jet impingement to improve condensation heat transfer has not been tackled in literature. Therefore, we present in this paper a first look at the problem. Initially, we pondered upon a sentence Lord Rayleigh wrote in 1911 about the generation of BF. `[as] the breath [was] led through a tube[, the] first deposit occurs very suddenly.'\cite{rayleigh1911} Upon performing a simple experiment of breathing through a paper straw against a mirror, we noticed the sudden appearance of a condensate spot. The spot had a shape similar to the straw exit, a circle of defined boundaries. However, to our surprise, the condensate spot was weakly influenced by the strength (speed) of our breath and the distance between the mirror and the straw exit. This led us to build a simple experimental setup to control the mentioned variables. We show here that condensate spots are manifestations of a recovery concentration concept. The recovery concentration concept is analogous to the recovery or adiabatic-wall temperature investigated by Hollworth and Wilson \cite{hollworth1983entrainment,hollworth1985entrainment}. In their work, they showed that consistent results were obtained upon basing Nusselt number correlations on the recovery temperature difference rather than the apparent temperature difference. In this work, we show that the recovery concentration manifests itself as a condensate spot which we call Breath Figure (BF) spot. This spot defines the effective area over which condensation of the jet's vapor takes place. Hence, we believe that quantifying this parameter is an essential step towards understanding condensation improvement by jet impingement.

\section{Experimental method}

\begin{figure}
\centering
\includegraphics[width=\textwidth]{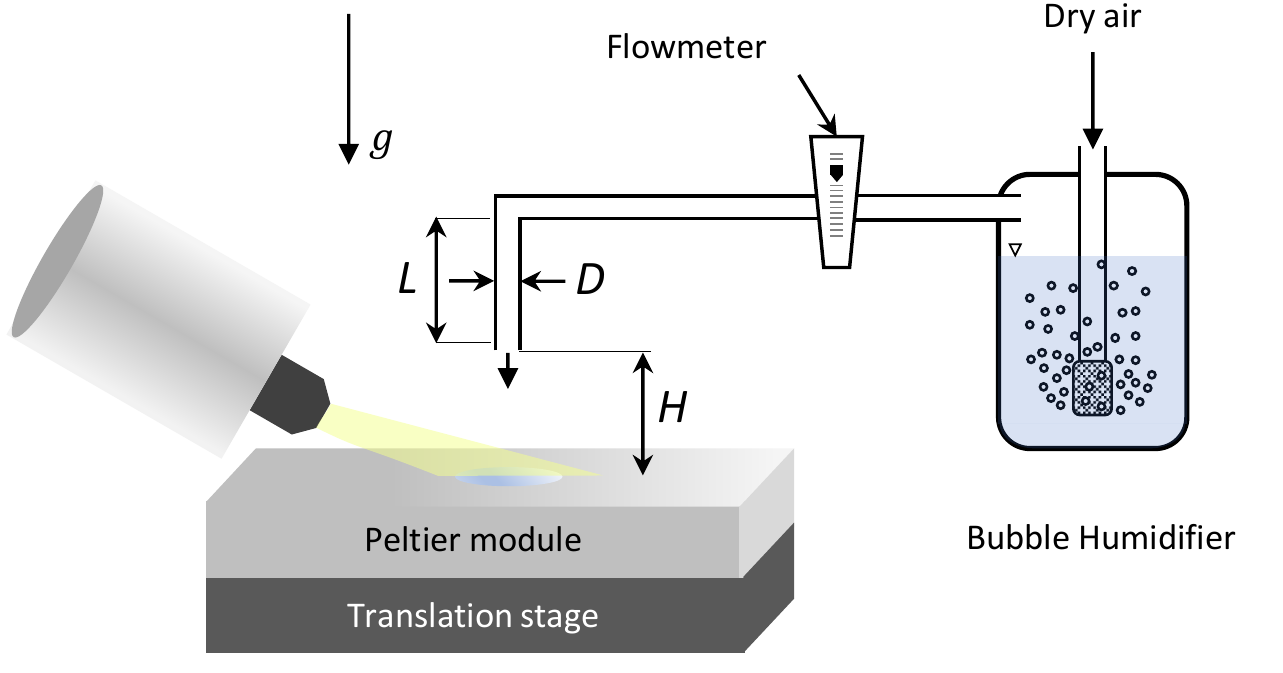}
\caption{Schematics of a table-top set-up for observing the BF spots from a jet of humid air under varying parameters namely jet-surface temperature difference ($T_j-T_s$), jet exit Reynolds number ($Re_j=v_jD/\nu$), and standoff-to-diameter ratio ($H/D$). }
\label{Fig1}
\end{figure}
\begin{figure}
\centering
\includegraphics[width=\textwidth]{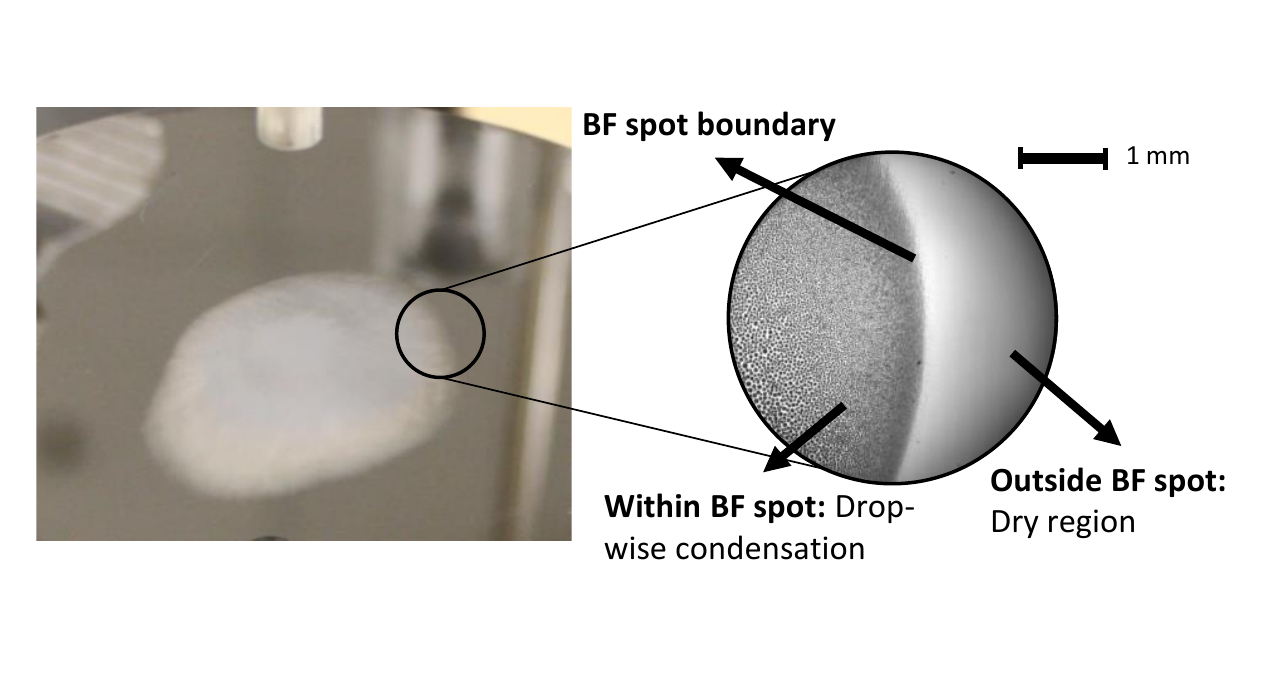}
\caption{Typical BF spot formation taken by a regular camera (left image) and a low magnification microscope (right image).}
\label{Fig2}
\end{figure}

In Figure \ref{Fig1}, we show the experimental setup which consists of a humidifier, a flow system, and a condensation surface. Dry air was first directed into a humidifier tank through several spargers to produce a humid air jet with the desired relative humidity. The humidifier tank was filled with DI water at room temperature, therefore, resulting in a room-temperature jet of humid air, $T_\infty=T_j$ = 22 $^o$C.  The flow rate of the air was controlled by a flow-adjustment valve and was measured using a rotameter (Walfront, model no. LZQ-7). Flow rate ranging from 1 LPM to 10 LPM were used in our experiments. The jet of humidified air exited a tube of diameter, $D$ = 3 mm, and a length, $L$ = 60 mm, that was located at a varying standoff distance, $H$ = 1 cm to 4.5 cm normal to the condensation surface.  The jet impinged normally on the surface in an ambient relative humidity of $RH_\infty$ = 20 \%. The jet exited the tube in a highly humid condition, $RH_j$ = 95 \%. This was achieved by placing three spargers (manufactured by Ferroday) to generate around 0.5 micron air bubbles in the humidifier tank, only one sparger is shown in Figure \ref{Fig1} for illustration. The condensation surface was an aluminum substrate that was placed on the cold side of a Peltier plate with a thermally conductive paste. The Peltier plate was supplied with an environmental chamber and a PID temperature controller (KR$\ddot{\rm{U}}$SS, DSA100). A range of substrate temperature, $T_s =$ 22 $^o$C to 5 $^o$C, was tested to observe the BF spot incipience and size variation. The temperature of the cold side of the Peltier plate was recorded using an RTD element that was supplied with the PID temperature controller (KR$\ddot{\rm{U}}$SS). An Infra-red (IR) camera (FLIR, A6753sc), and a flush-mounted k-type thermocouple (OMEGA, HH378) were used to observe the condensation substrate temperature as well as the condensate droplets. The substrate temperatures measured by the three methods were in agreement within 0.1 $^o$C. This rules out any possible heat transfer impedance of condensation due to surface thermal resistance. Systematic experiments were performed by first adjusting the flow to the desired jet Reynolds number $Re_j = 4 Q / \pi \nu D$, where $\nu$ is the kinematic viscosity of humid air. At the desired standoff-to-diameter ratio (H/D), the jet exiting the tube was allowed to impinge on the surface without lowering surface temperature initially. The surface temperature was then lowered in steps of 0.5 $^o$C from room temperature. At a certain surface temperature, we denote as the BF spot incipient temperature, BF spot starts to appear. As we lowered the surface temperature further, the expansion of the BF spot diameter was observed and recorded. The experimental parameters are summarized in Table 1 along with the colour/shape code of each data point. It is worth noting that a regular camera (Teledyne Photometrics, CoolSnap HQ2) was used to observe the BF spots. The camera was inclined with a maximum of 10$^o$ from the horizontal to obtain better visualization of the process. In Figure \ref{Fig2}, we show a typical BF spot observation from a selected experiment. The image on the left shows a macroscopic view of the BF spot while the right image shows a microscopic view (3X). Due to light scattering from the condensate micro-droplets, a white trace was observed upon looking at the condensate deposit. Under the microscope, the BF spot boundary becomes very distinct as it separate between a wet inside and a dry outside regions. Within the BF spot drop-wise condensation is observed as seen in Figure \ref{Fig2} (right) and Video 1. In Video 1 (supplementary material), we show a time lapse of the growth of sessile droplets near the BF spot boundary.

\begin{table}
\centering
\caption{Colour/shape code of the experimental conditions for a total of 35 combinations of $H/D$ and $Re_j$. Under each combination point, the temperature of the surface was varied from 22 $^o$C to 5 $^o$C and BF spot diameter was observed.\\}
\medskip
\begin{tabular}{c||c|c|c|c|c|c|c} \toprule
        & \textbf{$H/D$ =} 3.33 & 5 & 6.67 & 8.33 & 10 & 6.67 & 15\\ \hline
        \textbf{$Re_j$ =} 500 &$\medcircle$   & $\medblackcircle$     &  \color{blue} $\medcircle$ & \color{blue} $\medblackcircle$  & \color{red} $\medcircle$ & \color{red} $\medblackcircle$  & \color{Green} $\medblackcircle$ \\ \hline
        1340 &$\square$  & $\blacksquare$  &  \color{blue} $\square$ & \color{blue} $\blacksquare$ & \color{red} $\square$ & \color{red} $\blacksquare$ & \color{Green} $\blacksquare$\\ \hline
        2230 &$\triangle$   & $\blacktriangle$   &  \color{blue} $\triangle$ & \color{blue} $\blacktriangle$ & \color{red} $\triangle$ & \color{red} $\blacktriangle$ & \color{Green} $\blacktriangle$\\ \hline
        3130 &$\mdlgwhtdiamond$  & $\mdlgblkdiamond$  &  \color{blue} $\mdlgwhtdiamond$ & \color{blue} $\mdlgblkdiamond$ & \color{red} $\mdlgwhtdiamond$ & \color{red} $\mdlgblkdiamond$  & \color{Green} $\mdlgblkdiamond$ \\ \hline
        4130 &\FourStarOpen  & \FourStar &  \color{blue} \FourStarOpen & \color{blue} \FourStar & \color{red} \FourStarOpen & \color{red} \FourStar & \color{Green} \FourStar \\ \bottomrule
\end{tabular}
\end{table}


\section{Results and Discussion}

\begin{figure*}
\centering
\includegraphics[width=\textwidth]{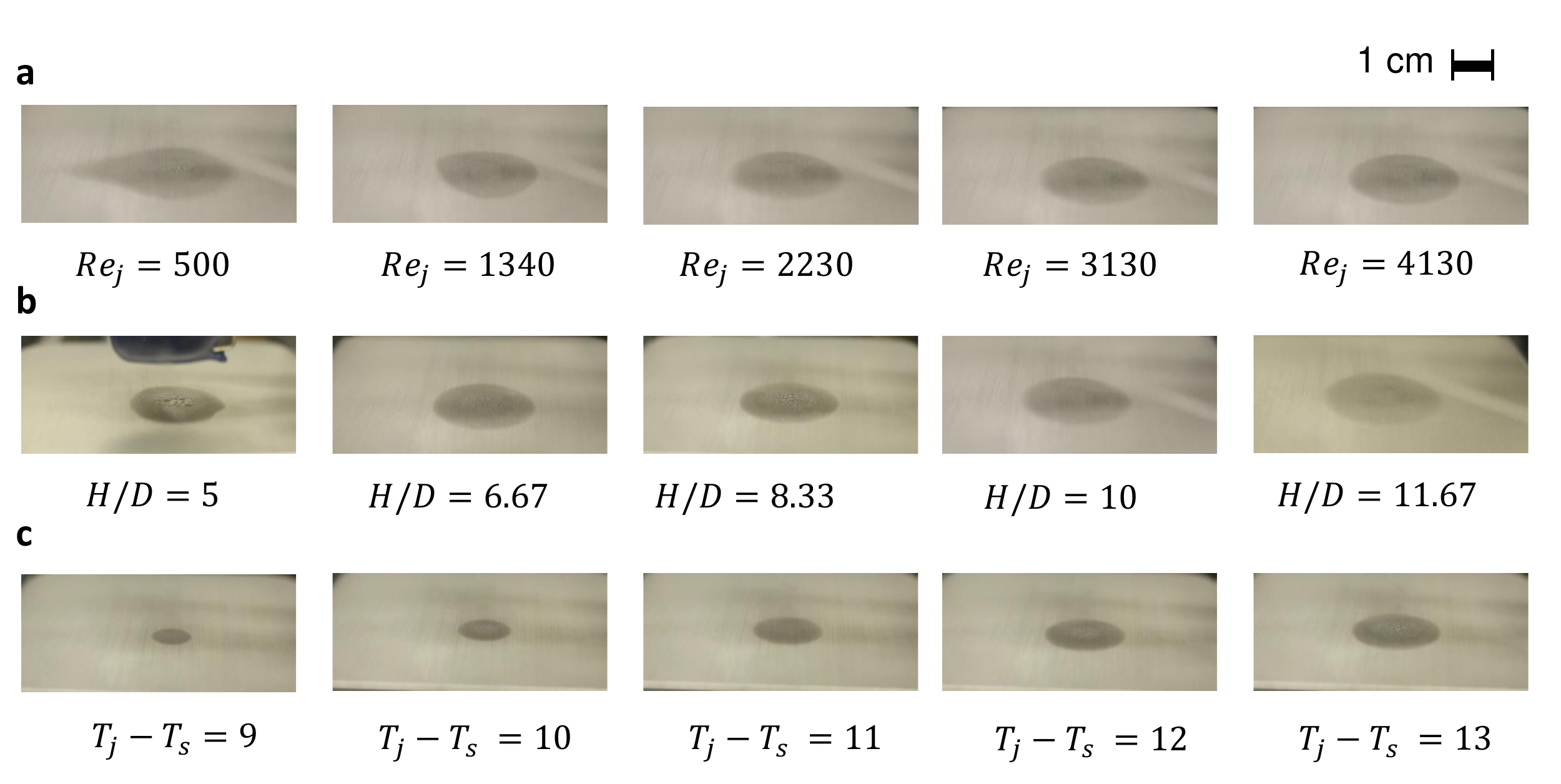}
\caption{Selected pictures of the BF spots at various conditions. \textbf{a.}  BF spots at varying jet Reynolds number. The selected pictures are for the case of $T_j-T_s$ = 18 $^o$C and $H/D$ = 10. \textbf{b.} BF spots at different standoff-to-diameter ratios. The selected pictures are for the case of $T_j-T_s$ = 18 $^o$C and $Re_j$ = 3130. \textbf{c.} BF spots at different jet-surface temperature differences. The selected pictures are for the case of $H/D$ = 8.33 and $Re_j$ = 3130.}
\label{Fig3}
\end{figure*}

Selected pictures at various experimental conditions are shown in Figure \ref{Fig3}(a-c). In Figure \ref{Fig3}(a), for fixed $T_j-T_s$ = 18 $^o$C and $H/D$ = 10, the effect of the Reynolds number is shown. We observed that at the lowest Reynolds number, $Re_j$ = 500, any obliqueness of the tube from the normal to the surface is characterized by a "tailed" BF spot. The tail is directed opposite to the angle of obliqueness. Adjusting the impingement angle to eliminate the tail served as an indication of a 90$^o$-angle impingement in our experiments. It should be noted that the tail is absent for higher Reynolds numbers for small inclination of the tube. It is worth mentioning that for $1000<Re_j<3000$ jets are in a transition regime whereas jets become fully turbulent for $Re_j>3000$ \cite{viskanta1993}. Therefore, The existence of a tailed BF spot might be due to the laminar behaviour of the jet. Additionally, the circularity of the BF spot is clear for high Reynolds number. The effect of standoff-to-diameter ratio for $T_j-T_s$ = 18 $^o$C and $Re_j$ = 3130 is depicted in Figure \ref{Fig3}(b). We observed that BF spot size is invariant with $H/D$ at least for the tested range of 3.33 to 15. In Figure \ref{Fig3}(c), we present the effect of jet-surface temperature difference for $Re_j$ = 3130 and $H/D$ = 8.33. As the temperature of the surface falls below the dew point of the jet center, the BF spot appears. The point of BF spot inception occurs at lower surface temperature as $H/D$ increases. Also, further decrease in the surface temperature corresponds to an increase in the BF spot diameter. The BF spot diameter keeps increasing with decreasing the surface temperature to the point at which atmospheric vapor start condensing. Below the atmospheric dew point, BF spot becomes indistinguishable from sessile droplets that appear on the entire surface.

\begin{figure}
\centering
\includegraphics[width=\textwidth]{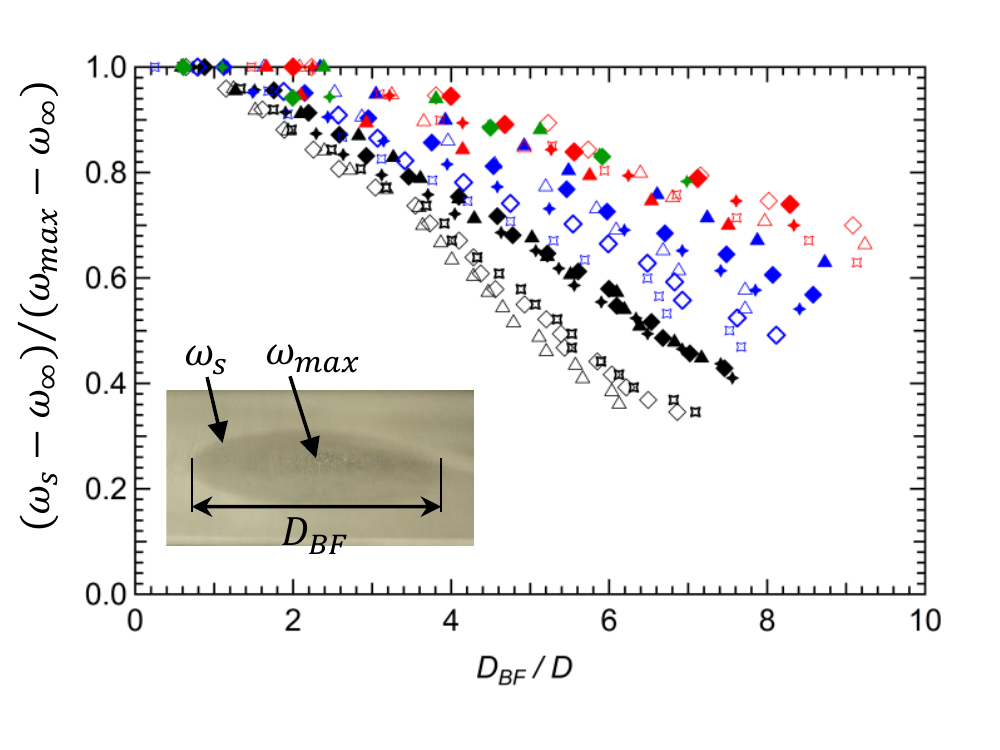}
\caption{Dimensionless concentration distribution on the surface as a function of dimensionless radial distance (or BF spot diameter to tube diameter ratio) ($D_{BF} / D$). Colour/shape code correspond to Table I. }
\label{Fig4}
\end{figure}

\begin{figure}
\centering
\includegraphics[width=\textwidth]{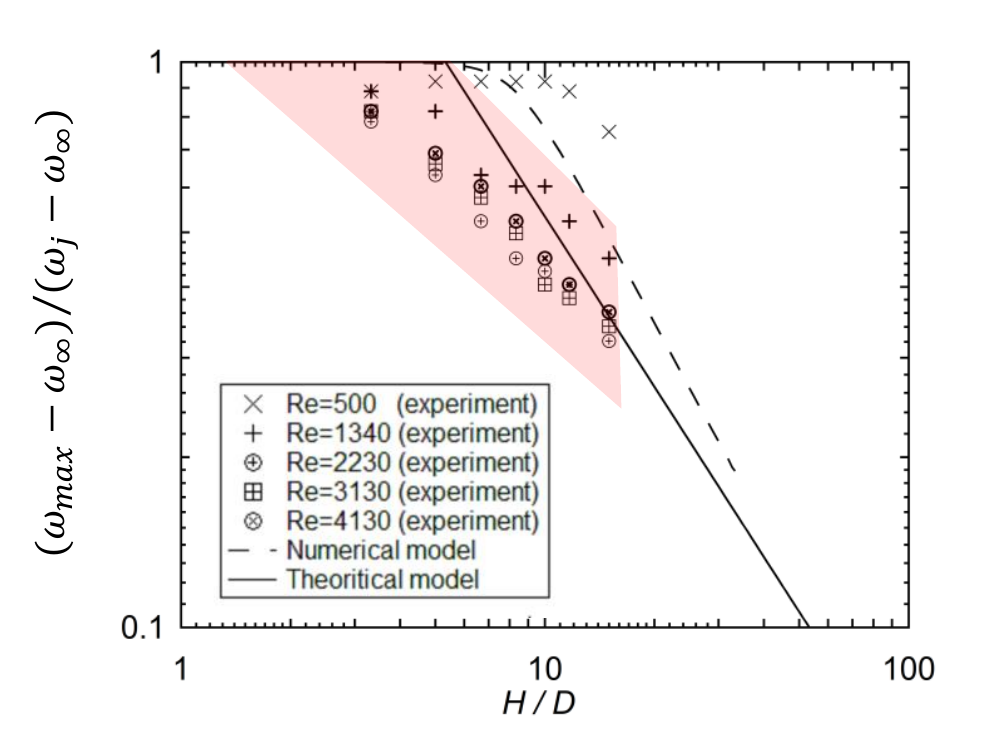}
\caption{Dimensionless maximum concentration as a function of $H/D$ and $Re_j$. The maximum concentration is obtained at the inception of BF spot point as depicted in Figure \ref{Fig4} and Eq. (\ref{Eq.1}). The red-shaded region correspond to the experimental uncertainty in measurements. }
\label{Fig5}
\end{figure}

To quantify that behaviour, vapor concentration distribution is inferred from the temperature measurements and BF spot size. At any experimental run, the vapor mass fraction at the boundaries of the BF spot was obtained as \cite{bird2006}
\begin{equation} \label{Eq.1}
\frac{1}{\omega_s} = 1 + \frac{M_a}{M_v}  (\frac{P_\infty}{P_v} - 1)
\end{equation}

where $M_a$, $M_v$, $P_\infty$, and $P_v$, are molecular weight of air, molecular weight of water, ambient pressure, and water vapor pressure at the surface temperature, respectively. In Figure \ref{Fig4}, we show the distribution of dimensionless vapor mass fraction $(\omega_j - \omega_\infty)/(\omega_{max} - \omega_\infty)$ as a function of normalized BF spot diameter $D_{BF}/D$ at different experimental conditions. The maximum mass fraction is obtained at the inception of BF spot. It should be noted that only results of turbulent jets ($Re_j>1000$) are shown in Figure \ref{Fig4} (The full range of Reynolds number is plotted in Figure S.1). First, It is clear that at any $H/D$, the mass fraction distribution is weakly influenced by Reynolds number. The lowest $H/D$ value shows a steeper drop of vapor mass fraction while increasing $H/D$ has a flattening effect. Further, for $H/D >$ 5, we observe that even standoff distance has a weak influence on the distribution of vapor mass fraction. It is worth noting that for $H/D < 5$, the free jet is still in the developing region \cite{Incropera2011}. Therefore, the behaviour becomes similar to a confined jet \cite{obot1982effect}. In Figure \ref{Fig5}, we plot the maximum vapor mass fraction as a function of both $H/D$ and $Re_j$. The vapor mass fraction is normalized with the jet excess mass fraction ($\omega_j - \omega_\infty$). We recognize that the laminar jet has a constant maximum vapor concentration for $H/D < 11.67$, which suggests that laminar jets lose less vapor content into the ambience compared to their turbulent counterparts. This is probably due to the improved mixing of the latter which helps in dissipating vapor to ambience. However, for turbulent jets, the maximum mass fractions seem to decrease monotonically with $H/D$ value.

The BF spot formation can be described using the following simple thought experiment. Consider a surface that is in thermal equilibrium with the jet and the ambience with the jet containing a higher vapor concentration than the ambience. After the humid jet exits the tube, vapor diffuses into the ambience before impinging on the surface. The concentration profile of the jet is therefore changed from being uniform to having a Gaussian distribution as shown in several analytical solutions \cite{gn1963theory,goodfellow2001industrial}. The value of the maximum concentration and width of the diffusing jet depends on the distance travelled by the jet as well as ambient thermal and flow conditions. Upon jet impingement, a significant mixing occurs that allows dissipation of the high vapor concentration near the surface. Without reducing the surface temperature, the vapor concentration at the wall has a decaying distribution from a maximum value at the center of the impingement area to a minimum value equivalent to that of the ambience further away. Due to the variation of the vapor mass fraction at the surface, there is an equivalent saturation temperature (dew point) variation with radial direction. For a constant surface temperature ($T_s$), When $T_s$ falls below the dew point at a given radial location, condensation will take place from the center up to that radial circumference, hence, a BF spot forms. 


In Supplementary material, we present a numerical model that employs our understanding of recovery concentration. The jet impingement on a wall is reduced to a two-dimensional axisymmetric problem. The jet issuing from the tube has fully developed velocity profile and uniform temperature, and concentration profiles. The continuity, momentum, energy and species equations are solved over the jet vicinity region. Because of the importance of accounting for turbulence in jet dynamics, standard $k-\omega$ formulation is usually preferred \cite{singh2017simulations,guide2006fluent}. A finite volume solver was utilized to obtain the solution of the governing equations in the desired domain. Solutions of the $Re_j$ and $H/D$ combinations were obtained both for free jet and wall obstructed cases. Further details and insights could be found in Supplementary material.

We first model theoretically the vapor concentration in the impingement region ($D_{BF}\leq5D$). Most importantly, we focus on the maximum concentration value which seems to be influenced significantly by the standoff distance rather than Reynolds number according to Figure \ref{Fig5}. To obtain a scaling analysis of such behaviour, we resort to the visual observations obtained from our numerical simulations in Figure S.3 and Figure S.4. The maximum concentration of a jet impinging on wall corresponds to its counterpart in a free unbounded jet at the same standoff-to-diameter ratio. In other words, the maximum vapor concentration is not affected by the impingement action. Therefore, we derive the theoretical curve in Figure \ref{Fig5} by using a free unbounded jet solution. Using a control volume at the tube exit to an arbitrary axial location in a free unbounded jet, momentum conservation can be written as
\begin{equation} \label{Eq.2}
    \rho v_j^2 \frac{\pi D^2}{4} = 2 \pi \rho \int_0^\infty v^2 r dr 
\end{equation}

where $\rho$ is the overall mixture density and $v_j$ is the mean velocity of the jet. By assuming that the overall mixture density does not vary greatly, which is valid for such low vapor concentrations. According to previous studies, in the developed region ($H/D >$ 5), the velocity profile has a Gaussian distribution form. Reichart gave the following relation at any given axial location (H) \cite{reichardt1942gesetzmassigkeiten,reichardt1944impuls}
\begin{equation} \label{Eq.3}
    v = v_{max} \exp[-a_v \bigg( \frac{r}{H} \bigg)^2]
\end{equation}

where $a_v$ is an empirical constant that depends on the tube exit geometry. Substituting Eq. (\ref{Eq.3}) into Eq. (\ref{Eq.2}), we get the maximum velocity as $v_{max}/v_j = \sqrt{a_v/2} (D/H)$. If we apply vapor species conservation over the same control volume, we have
\begin{equation} \label{Eq.4}
    \rho v_j (\omega_j - \omega_\infty) \frac{\pi D^2}{4} = 2 \pi \rho \int_0^\infty v (\omega - \omega_\infty) r dr 
\end{equation}

where $\omega_i$ is the vapor mass fraction evaluated at the surface temperature and saturated conditions, and $\omega_\infty$ is the vapor mass fraction evaluated at the ambient temperature and relative humidity. In general, the concentration profile of the jet has a Gaussian distribution as well. Therefore, one can write the concentration profile as
\begin{equation} \label{Eq.5}
    (\omega - \omega_\infty) = (\omega_{max} - \omega_\infty) \exp[- a_\omega \bigg( \frac{r}{H} \bigg)^2]
\end{equation}

where $a_\omega$ is an empirical constant different from that associated with the velocity profile. Substituting Eq. (\ref{Eq.3}) and Eq. (\ref{Eq.5}) into Eq. (\ref{Eq.4}) and combining the constants yield
\begin{equation} \label{Eq.6}
    \frac{\omega_{max} - \omega_\infty}{\omega_j - \omega_\infty} = \frac{a_v + a_\omega}{\sqrt{2 a_v}} \frac{D}{H}
\end{equation}

where the leading constant $(a_v + a_\omega)/\sqrt{2 a_v}$ is an empirical value that depends on the tube-exit type and experimental conditions. In table S.1, we present the experimental values of the leading constant for the different cases studied. Data of over 105 experiments show to be well represented by $(a_v + a_\omega)/\sqrt{2 a_v} = 5.3 \pm 2$. Eq. (\ref{Eq.6}) is depicted in Figure \ref{Fig5} along with the numerical model result. The theoretical model seem to capture maximum concentration behaviour within the experimental uncertainty. On the other hand, a small deviation is observed for the numerical simulation. Even though, the overall behaviour is captured by both methods, we believe that both methods have their limitations. The theoretical model assumes velocity and concentration to possess Gaussian distributions, however, several other profiles, such as a polynomial \cite{abramovich1984theory} could be used. The numerical simulation utilized the standard $k-\omega$ model which is highly sensitive to the inlet and boundary conditions. However, given the simplistic approach of predicting the general behaviour, both methods offer excellent predictive tools.  

Next, we use an analytical approach for ($D_{BF}/D > 5$) to analyse the BF spot boundary in the wall-jet region. Because of the sudden deposition, we can assume that BF spots are analogous to the concept of adiabatic wall or recovery temperature. Recovery temperature has been discussed in the context of heat transfer of impinging jets \cite{hollworth1983entrainment,hollworth1985entrainment,goldstein1986streamwise} as well as high Mach number flows \cite{eckertanalysis,dorrance2017viscous}. The importance of such parameter emerged from the mismatch between surface, jet and ambient temperatures which necessitates \textit{entrainment}. By the same token, we think BF spots are manifestations of a \textit{recovery concentration} concept that has not been discussed in literature as to the author's knowledge. Here we present a theoretical model of the recovery concentration. 

\begin{figure}
\centering
\includegraphics[width=\textwidth]{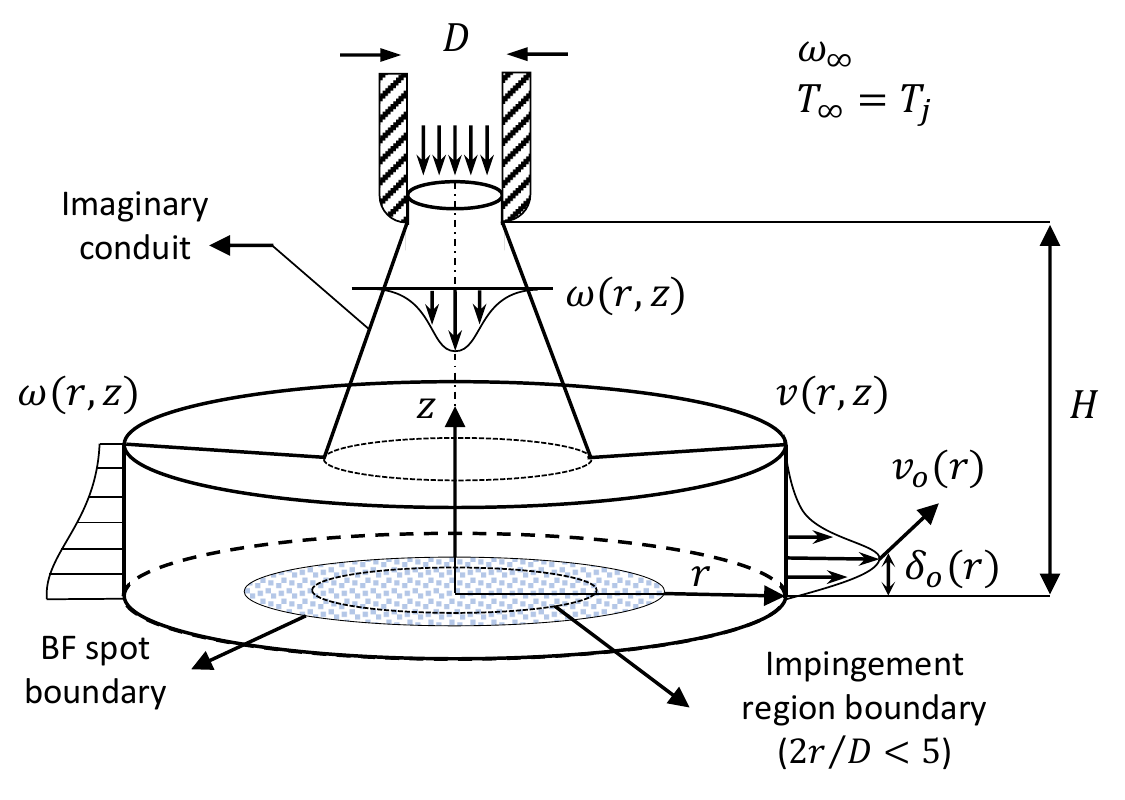}
\caption{Control volume approach for analysing humid air jet impingement. Schematic of the imaginary conduit over which vapor mass is conserved. Derivations of Eq. (\ref{Eq.6}) and Eq. (\ref{Eq.12}) depend on the understanding of this schematic. As the humid air exits the tube, vapor starts to diffuse into the ambience. However, the imaginary conduit boundary is located at a radial location where the gradient of vapor concentration is nearly zero, i.e. negligible diffusion is present. As the stream of vapor-air impinges on the surface, flow changes direction from y-direction to r-direction. The velocity and vapor concentration profiles at an arbitrary radial location away from the impingement region are depicted.
}
\label{Fig6}
\end{figure}

In Figure \ref{Fig6}, we present a schematic of an imaginary conduit starting from the tube exit and covering the impinged surface at an arbitrary radial location ($r$). At the bounding surfaces of the conduit, there is negligible mass transfer or negligible vapor mass concentration gradient. Applying a species mass conservation between the tube exit and the radial location on the surface gives
\begin{equation} \label{Eq.7}
    \rho \frac{\pi D^2}{4} v (\omega_j - \omega_\infty) = \rho \int_0^\delta v (\omega - \omega_\infty) (2 \pi r) dz
\end{equation}

where $\delta$ is the total thickness of the boundary layer. It has been recognized by several researchers that upon normalizing the velocity profile with its local maximum value, all velocity profiles in the wall jet region simplifies to $v/v_o \sim f(z/\delta)$ \cite{hollworth1983entrainment}. Whereas normalizing the excess local vapor mass fraction with the excess recovery concentration should result in a self-similar solution. Here we assume that, in the wall jet region, the non-dimensional vapor mass fraction is $\sim f(z/\delta)$. Upon performing the normalization, we obtain
\begin{equation} \label{Eq.8}
    r \delta (\omega_r - \omega_\infty) v_o \int_0^1 \bigg(\frac{v}{v_o}\bigg) \bigg(\frac{\omega - \omega_\infty}{\omega_r - \omega_\infty}\bigg) \frac{dz}{\delta} = \frac{D^2 v_j}{8} (\omega_j - \omega_\infty)
\end{equation}

For self-similar velocity and concentration profiles, the entire integral is assume to be a constant ($C_1$). Simplifying the previous relation gives
\begin{equation} \label{Eq.9}
    \frac{\omega_r - \omega_\infty}{\omega_j - \omega_\infty} = \frac{1}{8 C_1} \bigg(\frac{v_j}{v_o}\bigg)  \bigg(\frac{D}{r}\bigg) \bigg(\frac{D}{\delta}\bigg) 
\end{equation}

According to the several studies of turbulent jets \cite{glauert1956,poreh1967,hollworth1983entrainment}, the normalization thickness and velocity in the wall jet region can be correlated as
\begin{equation} \label{Eq.10}
    \frac{v_o}{v_j} = C_2 \bigg(\frac{H}{D}\bigg)^{0.1} \bigg(\frac{r}{D}\bigg)^{-1.1}  
\end{equation}
\begin{equation} \label{Eq.11}
    \frac{\delta}{D} = C_3  \bigg(\frac{r}{D}\bigg) 
\end{equation}

\begin{figure}
\centering
\includegraphics[width=\textwidth]{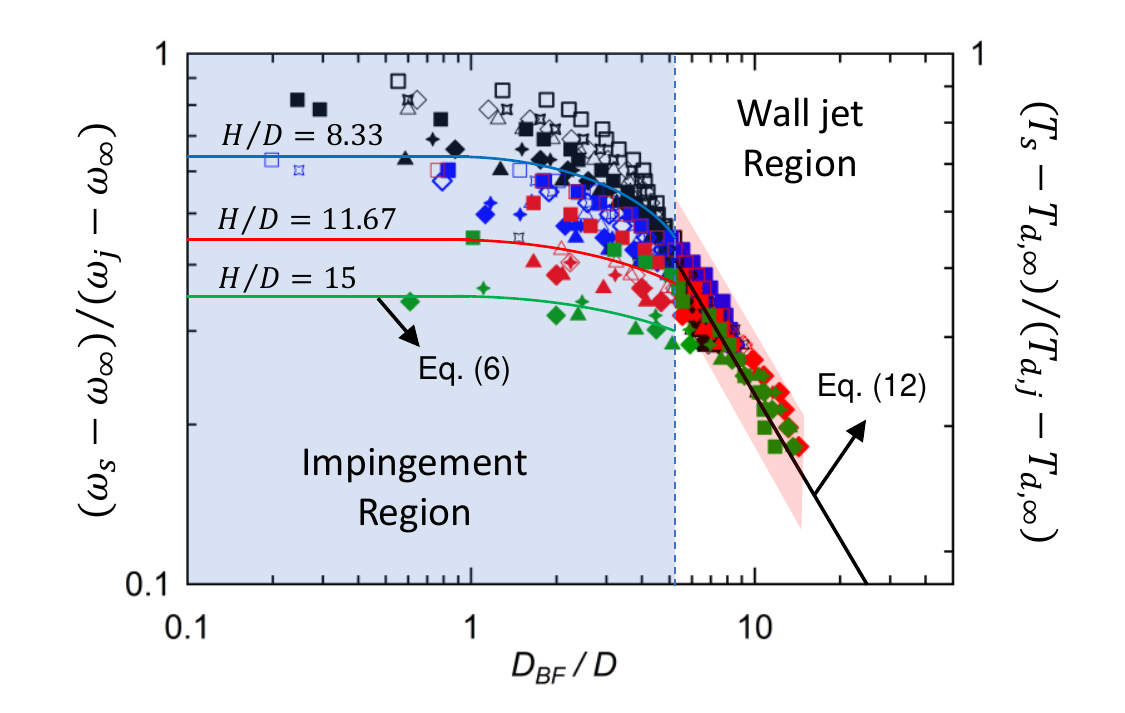}
\caption{Recovery concentration distribution. A plot of nondimensional vapor mass fraction and nondimensional surface dew temperature with respect to the extent of BF spot circle. The plot is split into two regions; impingement region ($D_{BF}/D < 5$ - blue-shaded region); and wall jet region ($D_{BF}/D > 5$). The derivation of Eq. (\ref{Eq.6}) is valid in the impingement region while Eq. (\ref{Eq.12}) is valid in the wall jet region. The red-shaded region correspond to the experimental uncertainty in measurements.}
\label{Fig7}
\end{figure}

Substituting Eq. (\ref{Eq.10}) and Eq. (\ref{Eq.11}) into Eq. (\ref{Eq.9}) and combining the constants result in the following conclusion
\begin{equation} \label{Eq.12}
    \frac{\omega_r - \omega_\infty}{\omega_j - \omega_\infty} = C_4 \bigg(\frac{H}{D}\bigg)^{-0.1} \bigg(\frac{r}{D}\bigg)^{-0.9}
\end{equation}

where $C_4 = 1/ (8 C_1 C_2 C_3)$ is a constant that depends on the tube-exit type and experimental conditions. Eq. \ref{Eq.12} shows a weak effect of standoff-to-diameter ratio with a power law of $-0.1$. Acceptable results within the experimental uncertainty could also be obtained if the effect of $H/D$ is absorbed into the leading constant. Figure \ref{Fig7} shows all the experimental data along with the theoretical curve given by Eq. (\ref{Eq.12}). The recovery concentration is independent of the jet Reynolds number at any given standoff-to-diameter ratio. Furthermore, there is no clear effect of standoff-to-diameter ratio in the wall-jet region. This is clear as all data points collapse on a universal curve in that region. The effect of standoff distance is noticed from Eq. (\ref{Eq.12}) to be very minimal which is in accord to our observation in Figure \ref{Fig3} and Figure \ref{Fig4}. Table S.1 presents the curve fitting constant obtained for the experimental data points. Data of over 1890 experiments show to be well represented by a leading constant ($C_4 = 1.12 \pm 0.14$) in Eq. (\ref{Eq.12}). We also showed mathematically that the jet velocity has no effect on the value of maximum vapor concentration. Eq. (\ref{Eq.6}) is depicted in Figure \ref{Fig7} where the effect of $H/D$ is pronounced at the center of the impingement region. The BF spot dimension between the center of the impingement region to the wall jet region varies smoothly in a transition region.

\section{Conclusion}

In conclusion, measurements were made of an isothermal humid air jet exiting a tube into a stagnant room-condition air. Because the humid air jet has higher vapor content than the environment, vapor diffuses from the former to the latter. It has been shown that humid laminar jets lose less vapor content as they travel into the environment compared to the turbulent jets because of the improved mixing mechanism of the latter. This was clear by observing the maximum concentration of the jet as it travel into an ambient air. On the other hand, for turbulent jets, the vapor content diffusion into the environment is independent of the jet's velocity magnitude. The maximum vapor concentration becomes a function of standoff distance only beyond the developing free jet region. We also showed for the first time that visible BF spots are manifestations of a new concept of a recovery concentration. We drew our analogy from the recovery temperature concept in heat transfer applications. The newly found concept is very important in studying species mass transfer due to jet impingement in general. Our findings show that BF spot is the area over which effective condensation takes place. Quantification of BF spot size is essential in optimizing the surface area of condensers as well as their temperatures to obtain effective condensation rates. We also predicted theoretically the concentration distribution on a surface exposed to humid air jet impingement. Eq. (\ref{Eq.6}) and Eq. (\ref{Eq.12}) present important conclusions with which concentration distributions on an impinged wall are found. We believe that this study is of great importance to optimize jet impingement heat and mass transfer rates. Several applications could utilize this work's findings, such as in textile drying, dehumidification technologies or exhaled breath condensate (EBC) technology \cite{horvath2005exhaled,hunt2007exhaled}.

\section*{Acknowledgment}
A. Alshehri would like to express his sincere gratitude to King Fahd University of Petroleum and Minerals (KFUPM), Dhahran, Saudi Arabia.

\bibliographystyle{unsrt}
\bibliography{bibliography.bib}
\newpage

\end{document}


\maketitle
\newpage

\beginsupplement

\begin{figure}
\centering
\includegraphics{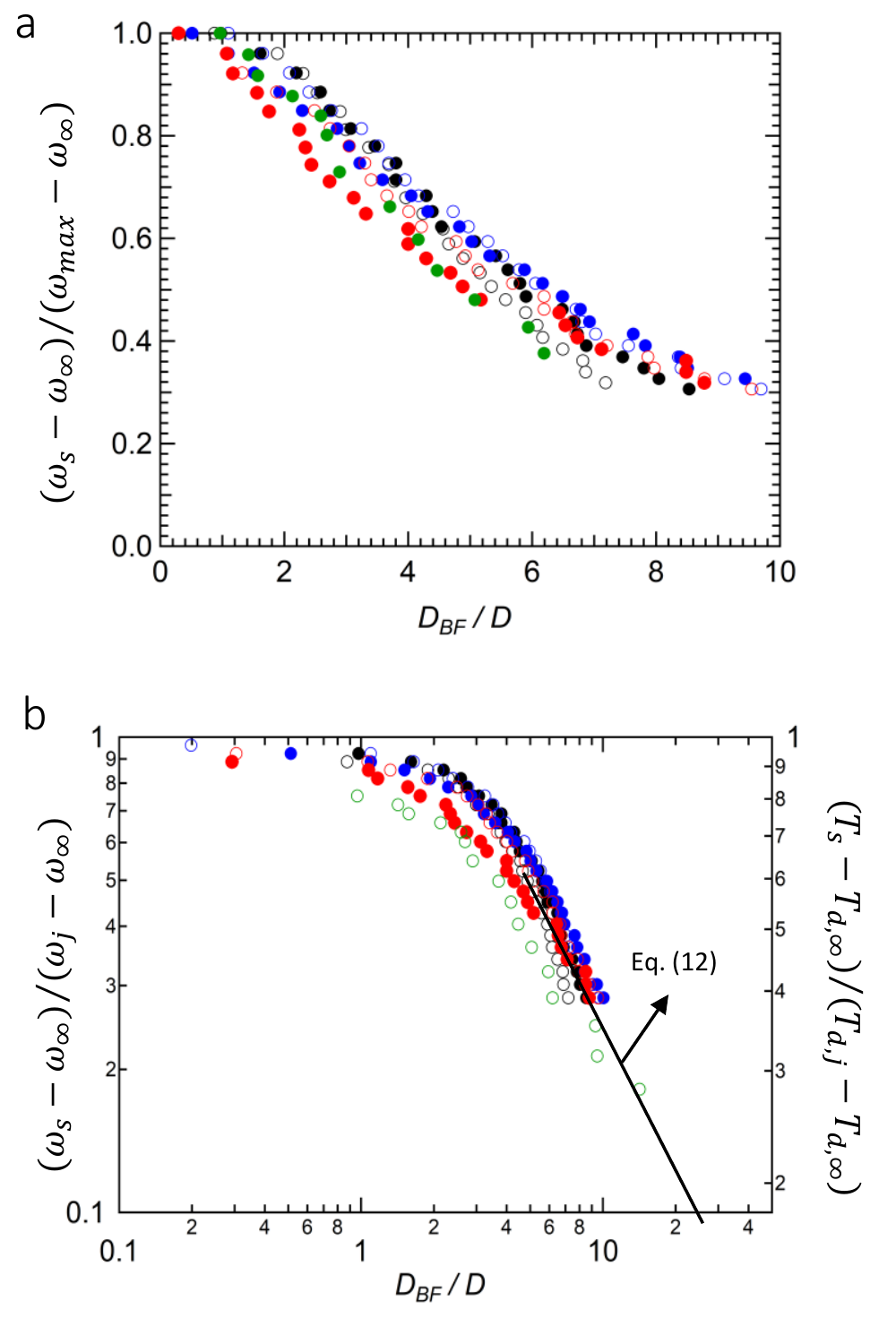}
\caption{Laminar jet experimental results. \textbf{a.}  Dimensionless concentration distribution on the surface as a function of dimensionless radial distance ($D_{BF} / D$). Colour and shape coding correspond to Figure 1(b). \textbf{b.} plot of nondimensional vapor mass fraction and surface dew-point temperature with respect to the extent of BF spot circle.}
\label{FigS1}
\end{figure}

\begin{table*}
\centering
\caption{Leading constant results from curve fitting of BF spot diameter, see equation (6) and equation (12). }
\medskip
\renewcommand{\arraystretch}{0.6}
\begin{tabular}{c c c c| c c c c}
\hline
$H/D$ & $Re_j$ & $(a_v + a_\omega)/\sqrt{2 a_v}$ & $C_4$ & $H/D$ & $Re_j$ & $(a_v + a_\omega)/\sqrt{2 a_v}$ & $C_4$ \\
\hline
3.33 & 500  & 2.97 & 1.08  & 10 & 500  & 9.23 & 1.38 \\
3.33 & 1340 & 2.97 & 1.04  & 10 & 1340 & 6.03 & 1.23 \\
3.33 & 2230 & 2.60 & 0.96  & 10 & 2230 & 4.27 & 1.11 \\
3.33 & 3120 & 2.73 & 1.04  & 10 & 3120 & 4.03 & 1.12 \\
3.33 & 4130 & 2.73 & 1.05  & 10 & 4130 & 4.5  & 1.18  \\ \hline
5 & 500  & 4.63 & 1.29   & 11.67 & 500   & 10.37 & 1.24\\
5 & 1340 & 4.10 & 0.99   & 11.67 & 1340  & 6.10 & 1.20\\
5 & 2230 & 3.17 & 1.03   & 11.67 & 2230  & 4.73 & 1.07\\
5 & 3120 & 3.30 & 1.08   & 11.67 & 3120  & 4.47 & 1.11\\
5 & 4130 & 3.47 & 1.10   & 11.67 & 4130  & 4.73 & 1.11\\
\hline
6.67 & 500  & 6.17 & 1.40 & 15 & 500   & 11.3 & 1.04 \\
6.67 & 1340 & 4.20 & 1.20 & 15 & 1340  & 6.73 & 1.21 \\
6.67 & 2230 & 3.50 & 1.11 & 15 & 2230  & 4.80 & 0.77 \\
6.67 & 3120 & 3.83 & 1.10 & 15 & 3120  & 5.10 & 0.92 \\
6.67 & 4130 & 4.03 & 1.10 & 15 & 4130  & 5.43 & 0.96 \\
\hline
8.33 & 500  & 7.70 & 1.46 \\
8.33 & 1340 & 5.03 & 1.27 \\
8.33 & 2230 & 3.73 & 1.14 \\
8.33 & 3120 & 4.13 & 1.11 \\
8.33 & 4130 & 4.37 & 1.11 \\
\hline\\
\end{tabular}
\label{table:Case1_Conditions}
\end{table*}

\newpage
\section*{ Numerical model development }

In this section, we utilize the finite volume method (FVM) to obtain the recovery concentration at various conditions. As was concluded in the paper, BF spots are manifestations of the recovery concentration concept. Therefore, impingement of humid air jet on an adiabatic surface is simulated. The geometrical domain as well as the boundary conditions are depicted in Fig. S.2a. The problem is reduced to an axisymmetric problem around an axis, at which no gradient in state variables is present in the radial direction. The jet originates from an inlet section of uniform velocity, temperature, and concentration profiles. The humid air flows through a tube of a length greater than the entry region to ensure fully developed conditions at the tube exit. The jet exits the tube into an ambient condition of given temperature, pressure and concentration preset to the outlet surfaces depicted in Figure S.2a. The impingement surface as well as the tube surface are characterized by zero heat and mass fluxes. The no-slip condition is applied to both surfaces as well. The flow of the humid air jet is assisted by gravitational force which acts normal to the impingement surface. The governing equations in the solution domain are given as
\begin{equation} \label{continuity}
\nabla \cdot (\rho \overrightarrow{v}) = 0
\end{equation}
\begin{equation} \label{Momentum}
\nabla \cdot (\rho \overrightarrow{v} \overrightarrow{v}) = -\nabla P + \nabla \cdot \overline{\overline\tau} + \rho \overrightarrow{g} 
\end{equation}
\begin{equation} \label{Energy}
\nabla \cdot (\overrightarrow{v} (\rho E + P)) = \nabla \cdot (k \nabla T - \sum_{j} h_j \overrightarrow{J_j})
\end{equation}
\begin{equation} \label{Species}
\nabla \cdot (\rho_j \overrightarrow{v}) = - \nabla \cdot \overrightarrow{J_j}
\end{equation}

where $E \approx h$ neglecting pressure work and kinetic energy. The total enthalpy is a mass weighted average of each species  enthalpy. The species enthalpy is given by equation (\ref{Enthaply}). 
\begin{equation} \label{Enthaply}
h_j =  \int_{T_{ref}}^{T} c_{p,j} dT
\end{equation}
The term $\overrightarrow{J_j}$ in equation (\ref{Energy}) and equation (\ref{Species}) refers to the diffusive mass flux of each species which is given by Fick's law.
\begin{equation} \label{Flux}
\overrightarrow{J_j} = -D_{j,i} \nabla \rho_j
\end{equation}

\begin{figure}
\centering
\includegraphics{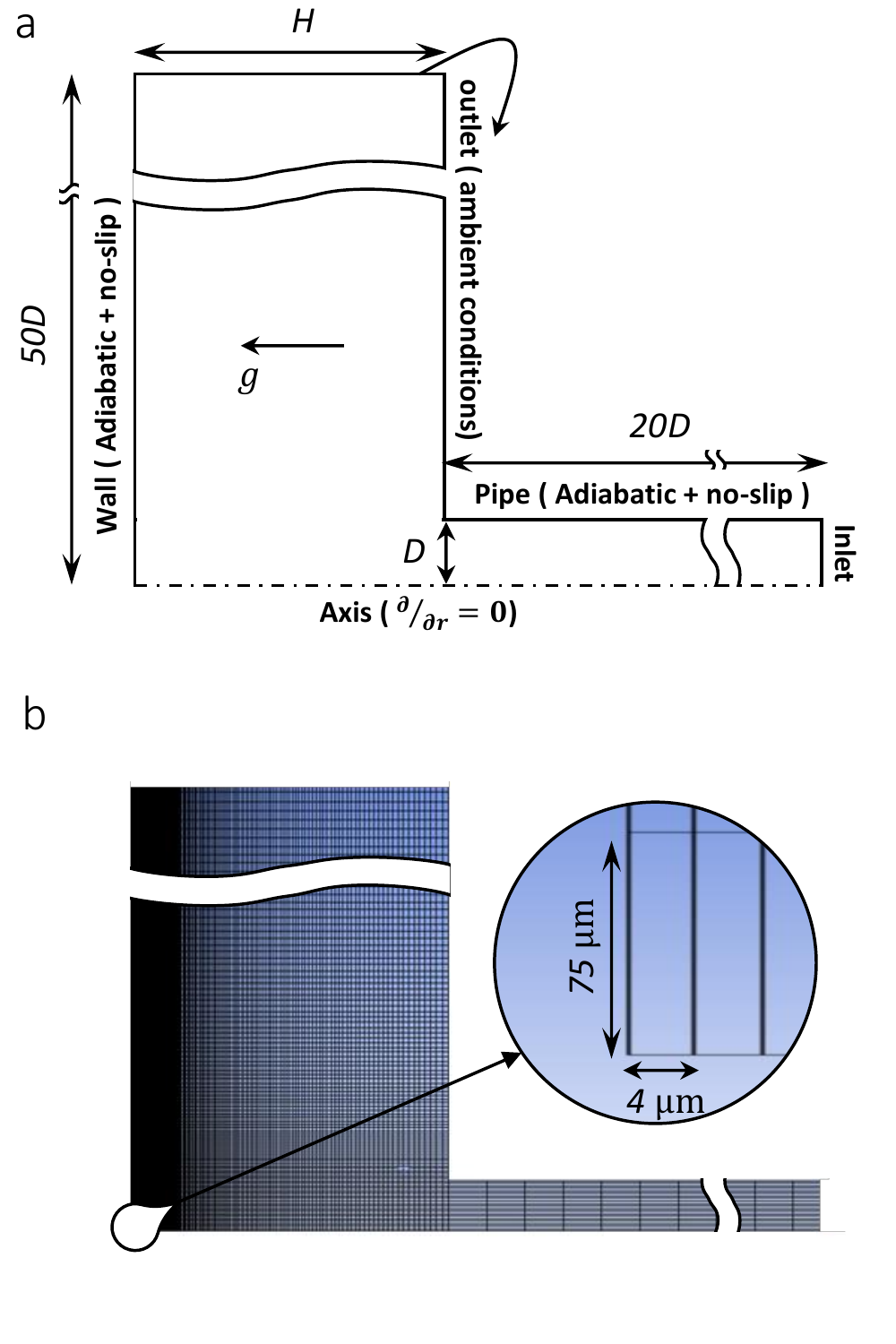}
\caption{Geometrical configuration of the numerical model. \textbf{a.} geometrical model of the axissymetric problem under simulation using FVM. \textbf{b.} Refined meshing of the solution domain.}
\label{FigS3}
\end{figure}

In order to take care of turbulence, standard $k-\omega$ model was implemented. Adding perturbed state variables to equations (\ref{continuity}-\ref{Species}), yields the extra term of Reynolds stress ($\rho \overline{u_i' u_j'}$). The standard $k-\omega$ model solves for two additional equations representing the transport of turbulent kinetic energy ($k$) and specific rate of dissipation ($\omega$). Enough documentation can be found in many references, therefore are not repeated here \cite{singh2017simulations,guide2006fluent}. The differential equations are solved using an FVM in which the domain is discretised into smaller cells as in Fig. S.2b. Finer meshing was concentrated where the change in state variables is expected to be greatest. It is worth noting that a separate simulation was performed on a free unbounded jet for comparison purposes.


Figure S.3 presents the contour plots of vapor mass fraction at varying standoff-to-diameter ratios while Fig. S.4 presents those of different Reynolds numbers. We observe that vapor concentration is maximum in the core of the jet. As the jet advances in the ambience, its vapor content diffuses and the uniform concentration tends to transition smoothly near the perimeter of the jet. Figure S.3 shows that for jets with heights that are 8.33 diameters or less, the maximum vapor concentration coincides with that of the inlet. The maximum vapor concentration then starts to drop due to the diffusion effect with ambience. From Fig. S.4, we notice that for Reynolds numbers of 1340 and higher, there is no significant difference of the vapor concentration profiles. The case of Reynolds of 500 shows slightly higher throw of vapor content. This could be attributed to the low mixing characteristic of laminar flows, therefore, maintaining its vapor content for a longer distance. For all the presented cases, the introduction of a wall normal to the jet flow direction does not seem to change the flow upstream. Therefore, further insights of the maximum vapor concentration at the wall could be obtained from a free jet case corresponding to similar geometric and flow conditions. This conclusion was the basis of the derivation of equation (6) in the main manuscript.

\begin{figure}
\centering
\includegraphics{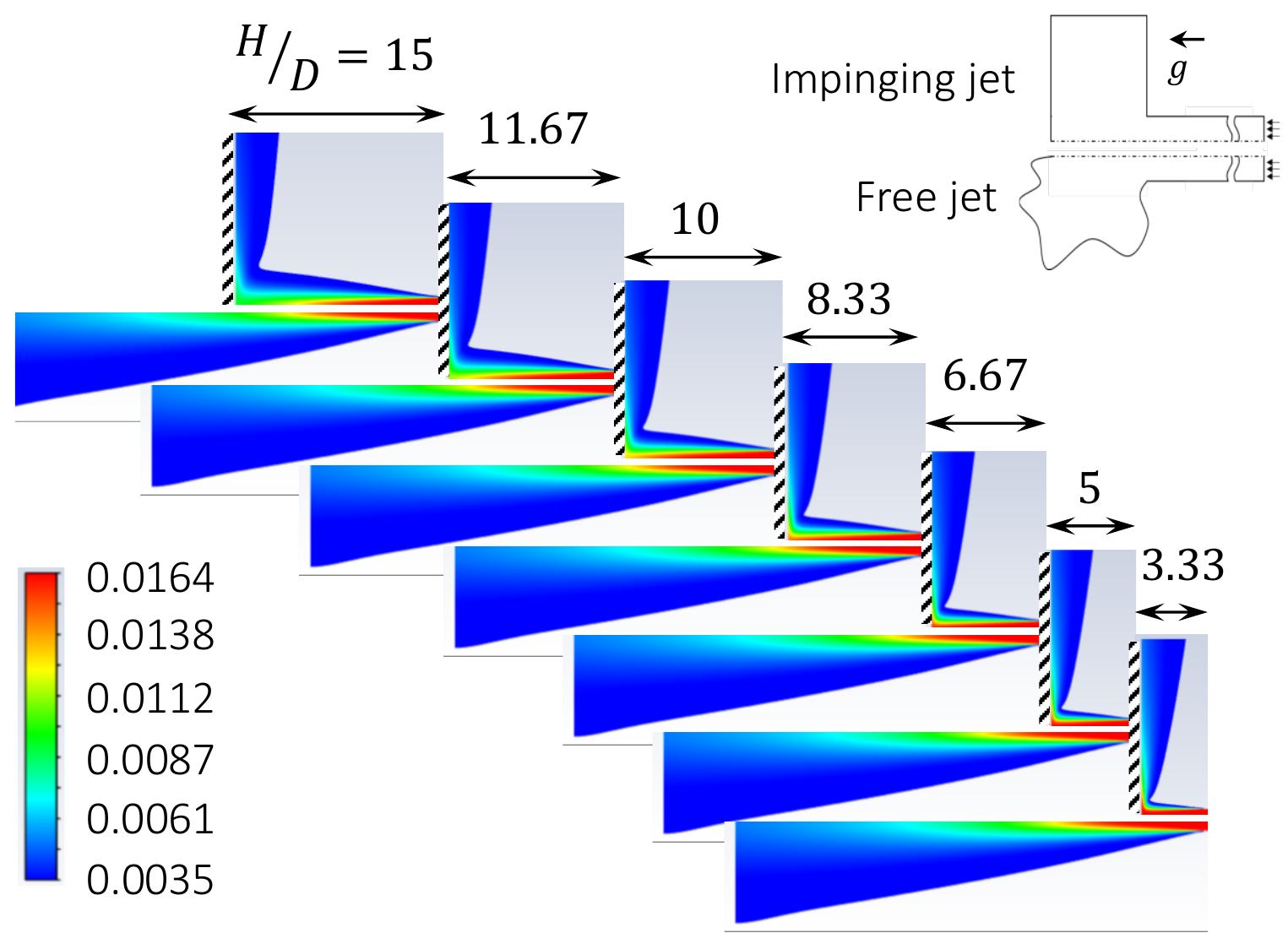}
\caption{Results of different standoff-to-diameter ratios. Contours of vapor mass fraction at $H/D$ of 3.33, 5, 6.67, 8.33, 10, 11.67, and 15 (from right to left). Results are for a selected Reynolds number of 4130. At each standoff-to-diameter ratio two cases are presented; (top contour plot) represents the case were a jet impinges on a wall corresponding to a given $H/D$; (bottom contour plot) represents the case of a free unbounded jet at a similar flow and geometric conditions. }
\label{FigS3}
\end{figure}

\begin{figure}
\centering
\includegraphics{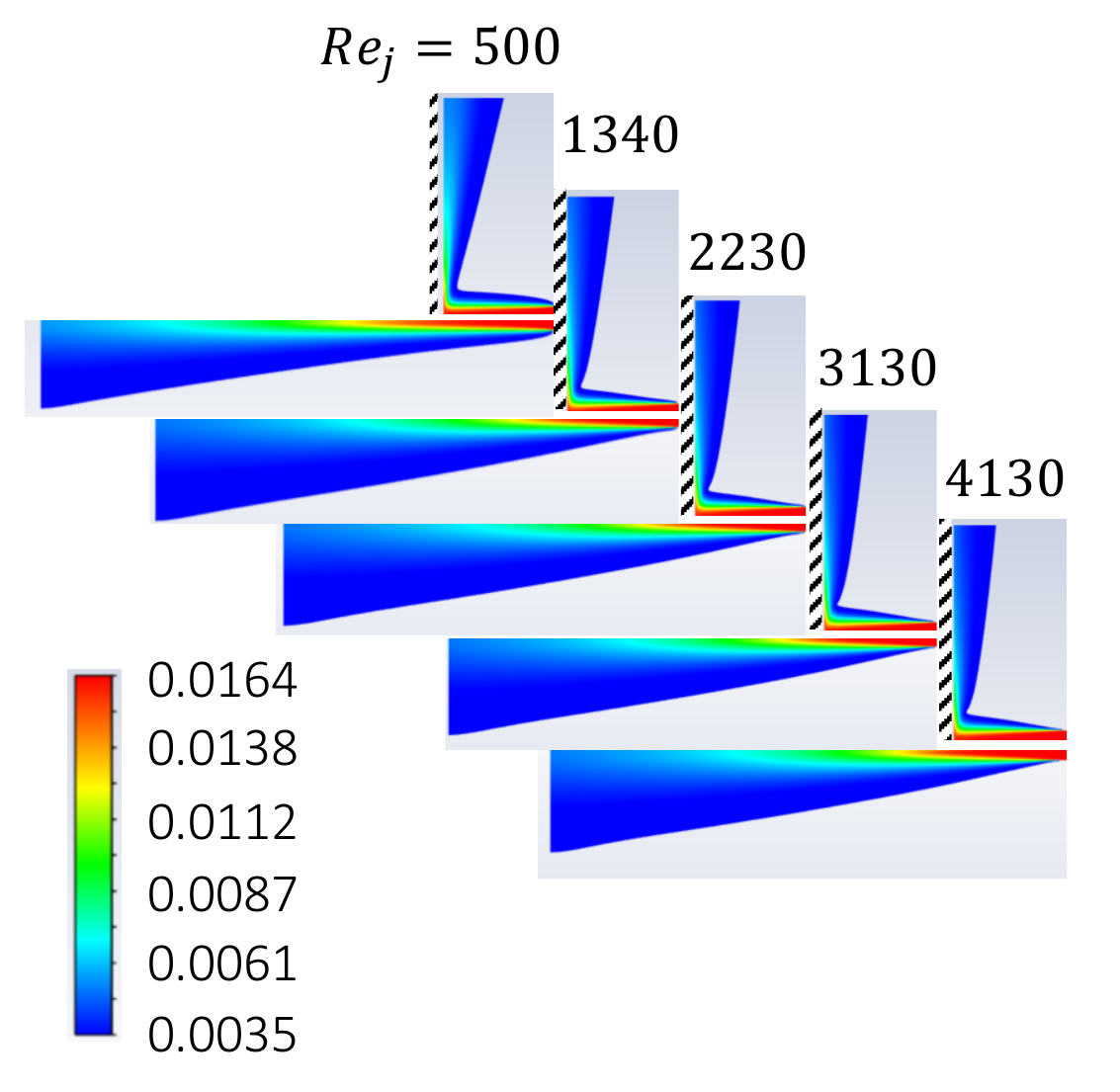}
\caption{Results of different Reynolds numbers. Contours of vapor mass fraction at $Re_j$ of 500, 1340, 2230, 3120, and 4130. Results are for a selected standoff-to-diameter ratio of 6.67. At each value of $Re_j$, two cases are presented; (top contour plot) represents the case were a jet impinges on a wall corresponding to the given $Re_j$; (bottom contour plot) represents the case of a free unbounded jet at a similar flow and geometric conditions. }
\label{FigS4}
\end{figure}

Figure S.5 depicts the vapor mass fraction normalized with the jet excess vapor mass fraction. We observe that within a radial location ($D_{BF}/D \leq 5$ (impingement region), the value corresponds to the maximum vapor mass fraction and is invariant with the radial location. Whereas for greater radial locations (wall jet region), there is almost a linear drop of the normalized vapor mass fraction. The effect of standoff-to-diameter ratio is negligible while the effect of Reynolds number is absent for the case of turbulent jets ($Re \geq 1340$). These conclusions are in line with our scaling analysis given in equation (6) and equation (13). Contrasted with Fig. 4, we observe a the obvious resemblance of the behaviour with a minor over-estimation by the numerical model.  

\begin{figure}
\centering
\includegraphics{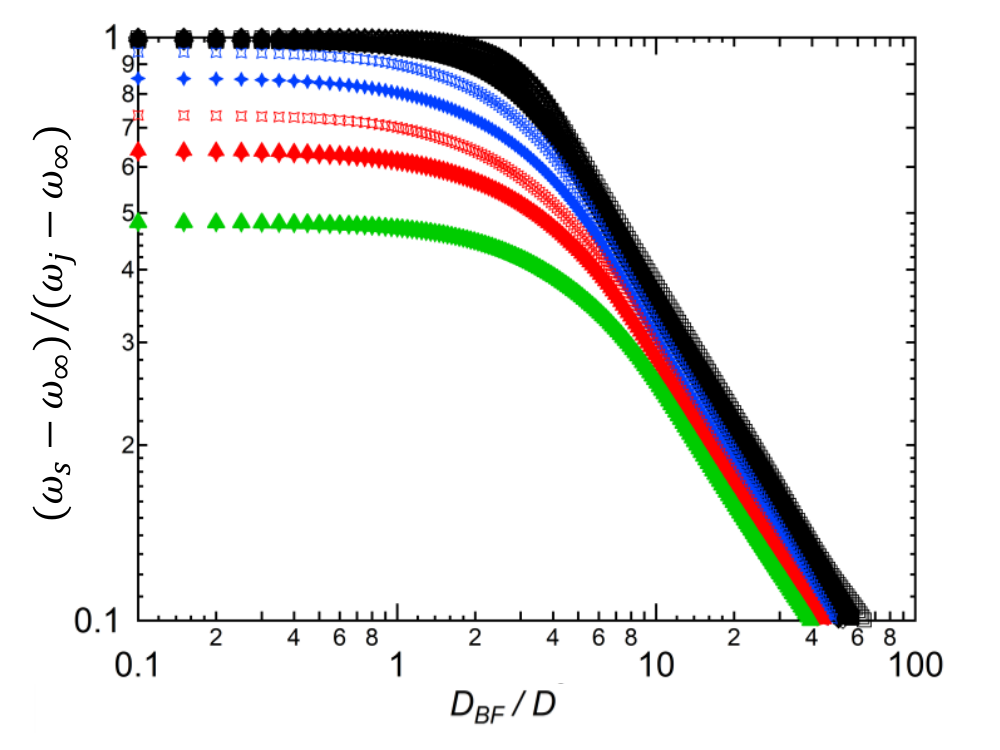}
\caption{Nondimensional recovery concentration. plot of nondimensional vapor mass fraction with respect to the extent of BF spot circle.}
\label{FigS5}
\end{figure}

\newpage

\bibliographystyle{unsrt}
\bibliography{bibliography.bib}
\newpage